# Increase in VEGF secretion from bone marrow stromal cell line (ST-2) induced by particles of porous silica glasses containing CuO and SrO

*Preethi Balasubramanian[1], Antonio J. Salinas[2,3], Sandra Sanchez-Salcedo[2,3], Rainer Detsch[1], Maria Vallet-Regi[2,3], Aldo R. Boccaccini[1*]*

[1]Institute of Biomaterials, University of Erlangen-Nuremberg, 91058 Erlangen, Germany

[2]Dpto. Química en Ciencias Farmacéuticas, Universidad Complutense, Instituto Hospital 12 de Octubre imas12, Madrid, Spain; [3]CIBER-BBN, Madrid, Spain.

***Corresponding author:** Aldo R. Boccaccini, E-mail: aldo.boccaccini@ww.uni-erlangen.de

Phone: +49(0)9131 85-28601

**Abstract**

Certain biomaterials are capable of inducing the secretion of Vascular Endothelial Growth Factor (VEGF) from cells exposed to their biochemical influence, which plays a vital role in stimulating angiogenesis. Looking for this capacity, in this study three porous glasses were synthesized and characterized. Glass compositions (in mol-%) were: $60SiO_2$–$(36-2x)CaO$–$4P_2O_5$ –$xCuO$–$xSrO$ with x= 0, 1 or 2.5, respectively, for B60, CuSr-1 or CuSr-2.5 glasses. $Cu^{2+}$ and $Sr^{2+}$ ions were added because the reported biological capabilities of $Cu^{2+}$ as angiogenic stimulator and $Sr^{2+}$ as osteogenic stimulator. The objective of this study was to determine the concentration of the glass particles that, being out of the cytotoxic range, could increase VEGF secretion. The viability of cultivated bone marrow stromal cells (ST-2) was assessed. The samples were examined with light microscopy (LM) after the histochemical staining for haematoxylin and eosin (HE). The biological activity of glasses was evaluated in terms of the influence of the $Cu^{2+}$ and $Sr^{2+}$ ions on the cells. The dissolution products of CuSr-1 and CuSr-2.5 produced the highest secretion of VEGF from ST-2 cells after 48 h of incubation. The combination of $Cu^{2+}$ and $Sr^{2+}$ lays the foundation for engineering a bioactive glass than can lead to vascularized, functional bone tissue when used in bone regeneration applications.

1. Introduction

Neovascularization is a critical step in bone repair and regeneration as bone is a vascularized tissue which relies on the close interaction between blood vessels and bone cells [1]. Several signaling molecules are involved in angiogenesis such as the Vascular Endothelial Growth Factor (VEGF), the basic fibroblast growth factor (bFGF), and various members of the transforming growth factor beta (TGFβ) family. VEGF is a signaling molecule secreted by hypertrophic chondrocytes and induces angiogenesis from the perichondrium leading to the recruitment of osteoblasts, osteoclasts and haemotopoietic cells [2]. VEGF, as a single agent, stimulates osteogenesis by activating osteoblasts and angiogenesis by activating endothelial cells to support bone repair [3, 4].

In the last decade, studies have shown that certain compositions of bioactive glasses stimulate the production of angiogenic growth factors such as VEGF both *in vitro* and *in vivo* [5–8]. In the last years novel bioactive glasses containing mesopores, i.e. in the 2-10 nm range, which exhibit novel features not present in the classical melt-derived bioactive glasses, have been developed by sol-gel methods [9]. These glasses were first proposed in 1991 [10] and their ordered mesoporous variants, mesoporous bioactive glasses (MBGs), were first proposed in 2004 [11]. MBGs were further characterized by Wu and Chang [12] and previous work in the field has been reviewed by Izquierdo-Barba and Vallet-Regí [13]. MBGs exhibit high *in vitro* bioactivity and controllable drug delivery capability which can be exploited for various biomedical applications including cancer therapy. Compared with traditional sol-gel derived glasses, MBGs exhibit ordered arrangements of mesopores and a very narrow pore size distribution. These features make MBGs more appropriate for the design of drug delivery and stimulus-responsive systems [9, 13]. However, for their applications in bone regeneration, both families of bioactive glasses exhibit similar behavior after implantation.

On the other hand, the incorporation of therapeutic ions such as lithium, boron, copper, strontium, magnesium, zinc, silver and others in the composition of any of the three families of bioactive glasses (i. e. melt-derived glasses, sol-gel derived glasses or MBGs), has been shown to be beneficial in influencing several biological functions such as osteogenesis, angiogenesis and anti-bacterial activity [14–17]. In this sense, ions such as $Sr^{2+}$ improve osteoblast differentiation by coordinating with various osteoblastic genes and they also have an effect on osteoclastic activity [18–21]. In addition, copper ions have been known for decades to stimulate angiogenesis and enhance the development of blood vessels to construct a vascularized structure [7, 14, 22–25].

In this work, three new bioactive glasses in the $SiO_2$–CaO–$P_2O_5$ system were obtained by a wet chemistry based method. Glasses investigated were based on the composition 60%$SiO_2$–36%–CaO–4%$P_2O_5$ (in mol-%), denoted as B60 glass. Furthermore, two novel compositions by incorporating different concentrations of CuO and SrO in B60, at the expenses of CaO, were produced and characterized (see Table 1). There is limited previous work on Sr-Cu-containing MBGs based on the $SiO_2$-CaO binary system, e.g. without P [26]. The present paper focuses on the evaluation of the secretion of VEGF from mouse stromal cells (ST-2 cells) from bone marrow when cultured in the presence of the dissolution products of bioactive glasses. The final aim of this study was to determine the concentration of the glass particles (size <32 μm) that, being out of the cytotoxic range, could increase VEGF secretion under the mentioned in vitro conditions, highlighting the angiogenic effect of the glasses in relation to the presence of $Sr^{2+}$ and $Cu^{2+}$ ions.

**Table 1:** Composition (in mol%) of glasses investigated in this study

| Acronym | SiO₂ | CaO | P₂O₅ | CuO | SrO |
|---|---|---|---|---|---|
| **B60** | 60 | 36 | 4 | - | - |
| **CuSr-1** | 60 | 34 | 4 | 1 | 1 |
| **CuSr-2.5** | 60 | 31 | 4 | 2.5 | 2.5 |

## 2. Experimental procedure

*2.1 Glass synthesis*

Glasses were synthesized by using the evaporation-induced self-assembly (EISA) method [27]. The nonionic surfactant Pluronic® _ P123 (Sigma-Aldrich) was used as a structure-directing agent. Pluronic® is an amphiphilic triblock copolymer having the sequence poly(ethylene oxide)$_{20}$–poly(propylene oxide)$_{70}$–poly(ethylene oxide)$_{20}$. During synthesis, 4.5 g of Pluronic® P123 was dissolved (1 h) in 85 mL of ethanol with 1.12 mL of 0.5 N $HNO_3$. Every 3 h interval other reactants were added under continuous stirring in the order: tetraethyl orthosilicate

(TEOS), triethyl phosphate (TEP), calcium nitrate, cooper nitrate and strontium nitrate (all the reactants of Sigma-Aldrich), keeping the flask covered by a plastic paraffin film (Parafilm®) in the amounts indicated in Table 2. Following the EISA process, the sol was cast in a Petri dish for gelation, which took 35 h. Then, gels were aged for 7 d at room temperature. The dried gels were removed as a homogeneous, transparent membrane and heated at 700 ºC for 3 h to remove the surfactant and nitrate groups, and to stabilize the resultant porous glasses under atmospheric conditions. Obtained materials were milled and sieved to obtain grains of size <32 μm giving rise to B60, CuSr-1 and CuSr-2.5 samples.

**Table 2:** Amount of reactants used in the glass synthesis and composition of the resultant glasses (in wt%) determined by EDX.

| Sample | TEOS (mL) | TEP (mL) | Ca(NO$_3$)$_2$·4H$_2$O (g) | Cu(NO$_3$)$_2$·2.5H$_2$O (g) | Sr(NO$_3$)$_2$ (g) | SiO$_2$ % | CaO % | P$_2$O$_5$ % | CuO % | SrO % |
|---|---|---|---|---|---|---|---|---|---|---|
| B60 | 7.67 | 0.79 | 4.54 | | | 62 ±2 | 34 ±1 | 4 ±1 | — | — |
| CuSr-1 | 7.67 | 0.79 | 4.14 | 0.13 | 0.12 | 63±2 | 30 ±3 | 4 ±1 | 0.8±0.1 | 1.6±0.7 |
| CuSr-2.5 | 7.67 | 0.79 | 3.47 | 0.33 | 0.30 | 62 ±5 | 32 ±5 | 2 ±1 | 2.3 ±0.3 | 1.5 ±0.5 |

*2.2 Glass characterization*

The mesopore structure was evaluated by transmission electron microscopy (TEM) in a JEM-2100 microscope (JEOL), operating at 200 kV, equipped with an energy dispersive X-ray (EDX; Oxford INCA) microanalysis stage. Nitrogen adsorption–desorption isotherms at 77.35 K used to determine the textural properties were acquired using a ASAP 2020 porosimeter (Micromeritics). Before the adsorption measurements, the samples were degassed under vacuum for 24 h at 120 ºC. The surface area was obtained by applying the Brunauer–Emmett–Teller (BET) method. The pore size distribution was determined by the Barrett–Joyner–Halenda (BJH) method from the adsorption branch of the isotherm [28]. Magic angle- spinning (MAS) and single-pulse solid-state nuclear magnetic resonance (NMR) measurements were performed to evaluate the different silicon and phosphorus environments in the synthesized samples. The NMR spectra were recorded on an Advance 400 spectrometer (Bruker). Samples were spun at 10 kHz for $^{29}$Si and 6 kHz for $^{31}$P. Spectrometer frequencies were set to 79.49 and 161.97 MHz for $^{29}$Si and $^{31}$P, respectively. Chemical shift values were referenced to tetramethylsilane and $H_3PO_4$ for $^{29}Si$ and $^{31}P$, respectively. All spectra were obtained using a proton enhanced cross polarization (CP) method, using a contact time of 1 ms. The time periods between successive accumulations were 5 and 4 s for $^{29}Si$ and $^{31}P$, respectively, and the number of scans was 10.000 for all the spectra.

*2.3 Cell culture*

Cell culture experiments were performed according to the previously described procedure [29–31]. A bone marrow stromal cell line (ST-2, Deutsche Sammlung für Mikroorganismen und Zellkultur, Germany), isolated from bone marrow of BC8 mice, was used for cell culture experiments. Cells were cultured in RPMI 1640 medium (Gibco, Germany) containing 10 vol% FBS (Sigma-Aldrich, Germany) and 1 vol% penicillin/streptomycin (Sigma-Aldrich).

ST-2 cells were seeded for 24 h at 100.000 cells/mL. At the same time, 0.1 g of the three glass powders were added to 10 mL culture media (without cells) and incubated separately for 24 h at 37 °C. After 24 h, the supernatant was extracted and diluted into different concentrations (0; 0.01; 0.1; 1 wt/vol) from all samples. The seeded ST-2 cells were washed with Phosphate Buffered Saline (PBS) and the different dilutions of supernatant from the pre-incubated granules were transferred to the ST-2 cells for further 48 h. In these tests, the cultured cells do not come in direct contact with the glass granules, but only with the ionic dissolution products.

A WST-8 assay was carried out to evaluate the viability of the cultivated cells, as described previously [29]. The samples were examined with light microscope (LM) after the histochemical staining using haematoxylin and eosin (HE).

*2.4 VEGF release*

The amount of released VEGF from ST-2 cells into the cell culture medium was measured by using a RayBio Human VEGF ELISA (Enzyme-Linked Immunosorbent Assay) kit. This assay, for the quantitative measurement of VEGF in cell culture supernatants, employs an antibody specific for mouse VEGF coated on a 96-well plate. The changes of color from blue to yellow are detected and the intensity of the color is measured at 450 nm (Phomo, Anthos Mikrosysteme GmbH, Germany). The assay procedure was performed according to the manufacturer's instructions.

# 3  Results

## 3.1 Glass characterization

Figure 1 shows TEM images and EDX spectra of the three synthesized glasses B60, CuSr-1 and CuSr-2.5. TEM micrographs show the absence of a mesoporosus order, although in some areas a worm-like mesopore structure can be seen. On the other hand, EDX confirms the presence of Si, Ca, P, Sr and Cu in the glasses in analogous amounts to the nominal glass composition. The elemental weight-percentages obtained are included in Table 2. The presence of Ni detected by EDX comes from the Ni-grids used in these studies (because the Cu content in two of the investigated glasses).

Nitrogen adsorption/desorption isotherms and pore size distribution curves of B60, CuSr-1 and CuSr-2.5 are shown in Figure 2. Inset tables display the specific surface area ($S_{BET}$), pore diameter ($D_P$) and pore volume ($V_P$) values for each sample. Isotherms can be identified as type IV according to the IUPAC classification, typical of mesoporous solids [32]. Moreover, the presence of H1 type hysteresis loops in the mesopore range indicated the existence of open ended cylindrical mesopores with narrow pore size distributions [28, 33]. As observed, $S_{BET}$ values ranged between 155 m$^2$/g (CuSr-2.5) and 210 m$^2$/g (CuSr-1) being 174 m$^2$/g for B60. Such surface areas are of the same order of magnitude of those of traditional sol-gel derived glasses and somewhat lower than that of MBGs. On the other hand, the highest values of $D_P$ and $V_P$ were measured for B60 (6.0 nm and 0.40 cm$^3$/g, respectively) whereas for CuSr-1 and CuSr-2.5 $D_P$ and $V_P$ were close to 4.0 nm and 0.30 cm$^3$/g.

$^{29}$Si solid-state MAS NMR measurements were performed to investigate the environments of the network former species at the atomic level as shown in Figure 3. $Q^2$, $Q^3$, and $Q^4$ represent the silicon atoms denoted as Si* in (NBO)$_2$Si*–(OSi)$_2$, (NBO)Si*–(OSi)$_3$ and Si*(OSi)$_4$ (NBO= non-bonding oxygen), respectively. The chemical shifts and the deconvoluted peak areas for each sample are shown. The signals in the -110 ppm region come from $Q^4$ and those at -100 ppm come from $Q^3$. A resonance at approximately -92 ppm comes from $Q^2$, but for CuSr-1 and CuSr-2.5 samples, the resonance decreases to -87 ppm [34] due to the presence of Cu$^{2+}$ and Sr$^{2+}$ in the network.

$^{29}$Si-MAS-NMR spectroscopy was used to evaluate the network connectivity (NC) of porous glasses as a function of chemical composition ($Q^n$) (Figure 3). B60 sample is characterized by a high percentage of $Q^4$ and $Q^3$ species and a NC of 3.20. The introduction of CuO and SrO causes a slight decrease of NC relative to B60 sample, increasing the percentage of $Q^2$ in detriment of $Q^3$. Therefore, when the concentration of CuO and SrO is increased in the glass, the NC decreased to 3.08 and 2.98 for CuSr-1 and CuSr-2.5 glasses, respectively. These samples exhibit a higher amount of ion modifier that disrupts the mesophase formation, leading to a depolymerized silica network [35]. This is because the amount of $Ca^{2+}$ necessary to act as a compensator of charge is higher than that present in the sample because some of the $Ca^{2+}$ ions are involved in the $Q^0$ species of P. This is indicative that the joint presence of $Ca^{2+}$ and $PO_4^{3-}$, results in amorphous calcium phosphate clusters located at the pore walls surface favoring the solubility of these materials, in agreement with a previously reported model [36, 37].

In addition, $^{31}$P-NMR spectroscopy was used to evaluate the local environment of P atoms, thus elucidating the phosphate species contained in the different samples (Figure 4). $Q^0$ and $Q^1$ represent phosphorus atoms (denoted by P*) in the $PO_4^{3-}$ species of P*–(NBO)$_4$ and (NBO)$_3$–P*–(OP), respectively. The spectra recorded by single pulse show a signal at ~2 ppm corresponding to $Q^0$ units present in the $PO_4^{3-}$ species [38]. The Full Width at Half Maximum (FWHM) for CuSr-2.5 and CuSr-1 samples are narrower than B60 indicating that the clusters are larger and more crystalline with increasing content of $Cu^{2+}$ and $Sr^{2+}$ ions. These results evidence that most of the P atoms are included as independent $PO_4^{3-}$ tetrahedra within the silica network, thus avoiding polyphosphate formation. The orthophosphates would be balanced with the divalent ions introduced in the system and the nature of these cations seems to play an important role on the characteristics of the phosphate clusters. Thus, $Cu^{2+}$ and $Sr^{2+}$ do not act as network formers but as network modifiers [39].

*3.2 Cell culture*

Cell viability values of ST-2 cells cultured with the supernatant of B60, CuSr-1 and CSr-2.5 glasses of different sizes and at different concentrations (0.01–1 wt/vol %) after 48 h of incubation are shown in Figure 5. Compared to the reference, B60 samples show a slight reduction in cell viability whereas the 0.1 and 1 wt/vol% of Cu- and Sr- containing glasses were cytotoxic. There is a significant reduction of cell activity through the released products from the

CuSr-1 and CuSr-2.5 particles. Similar results were obtained from LM images, as shown in Figure 6. The three dilutions of B60 and the 0.01 wt/vol% dilution of CuSr-1 and CuSr-2.5 have no negative influence on cell morphology, whereas higher concentrations, such 0.1 and 1 wt/vol%, were seen to disrupt the cell layer formation.

*3.3 VEGF release*

In Figure 6, the VEGF release from ST-2 cells cultured in the supernatant of different dilutions of B60, CuSr-1 and CuSr-2.5 after 48 h is shown. Similar to the cell viability results and LM images, the different dilutions of B60 glasses showed comparable VEGF release as the reference. Interestingly, the VEGF secretion from 0.01 wt/vol% dilution of CuSr-1 and CuSr-2.5 was significantly higher than the values of the reference and the B60 sample. The other two dilutions (1 and 0.1 wt/vol%) of CuSr-1 and CuSr-2.5 showed a decline in VEGF release.

# 4 Discussion

The EISA method allowed to readily prepare two porous glasses based in the mesoporous 60%$SiO_2$-36%CaO-4% $P_2O_5$ composition containing simultaneously CuO and SrO. First, including 1% of each extra element and secondly incorporating 2.5%. The glass characterization showed that $Cu^{2+}$ and $Sr^{2+}$ ions behave as network modifiers. In this way, they can exert their positive biological action, which is the reason why they were included in the glass. Moreover, Cu and Sr-containing glasses exhibited moderate textural properties and worm-like mesopores arrangement. However, BET surface areas close to 200 $m^2$/g and volumes of pores around 0.30 $cm^3$/g were high enough to permit significant surface reactivity when the glasses were investigated in cell culture.

In this sense, the synthesis of porous silica-based glasses doped with biologically active ions is gaining popularity among researchers [6, 7, 14, 15, 17–25, 35, 39, 40]. Copper is one such ions which has been known for more than two decades to play a significant role in angiogenesis [22–25, 41]. Rabbits with copper-deficiency were unable to produce an angiogenic response regardless of the type of angiogenic stimulus applied [42]. Notable amounts of cellular copper have been found in human endothelial cells when they were induced to undergo angiogenesis [41]. In recent years, copper has been incorporated into bioactive glasses in the form of scaffolds, particles and fibers for several biomedical applications [7, 17, 23–25, 43, 44]. On the other hand, strontium is considered as a promising agent in treating osteoporosis [18, 19, 21, 45, 46]. Moreover, Marie et al. reported the therapeutic capability of strontium in bone regeneration *in vivo* [47, 48] and a Sr-based drug called strontium ranelate has beneficial effects in bone healing [49]. In this study, different concentrations of Cu and Sr were added to the base BG composition $SiO_2$–$CaO$–$P_2O_5$ (Table 1) and new BG CuSr-1 and CuSr-2.5 were synthesized. The 1 and 0.1 wt/vol% concentrations of CuSr-1 and CuSr-2.5 were found to be cytotoxic for ST-2 cells. This could be due to the rapid release of copper and strontium ions into the solution, which ultimately increases the pH of the medium and kills the cells. However, the VEGF secretion was found to be the highest for the Cu and Sr containing glasses (0.01%). Since the cells were negatively affected by the release of ions, the 1 and 0.1 wt/vol% dilutions of CuSr-1 and CuSr-2.5 showed a decreased VEGF release. It is also important to note that the experiment was carried out in static conditions and considering the dynamics of the reactions occurring *in vivo*, these compositions of glasses could bring out favorable cell biology results.

As mentioned above, these elements (Cu, Sr) have been independently incorporated in other bioceramics. For instance Sr has been frequently included in calcium phosphates [50–52]. In this sense Sr-substituted hydroxyapatite materials were shown to promote the proliferation, osteogenic differentiation and angiogenic factor (VEGF) expression of human osteoblast-like cells [50]. Moreover, scaffolds based on calcium phosphate bioceramic containing Sr could apparently accelerate in vivo segmental bone regeneration through stimulating VEGF secretion from osteoblasts in rabbits [51]. Lin et al. further showed that Sr and Si ions released from Sr-substituted calcium silicate bioactive ceramic scaffolds acted synergistically to stimulate the osteogenic differentiation of MSCs and angiogenesis of umbilical vein endothelial cells in vitro, promoting bone regeneration and angiogenesis in a critical-sized calvarial defect model of ovariectomized rat [52].

On the other hand, several authors demonstrated that the release of copper ions stimulates the expression of proangiogenic factors such as VEGF and transforming growth factor-β (TGF-β) in wounds created in diabetic mice [53, 54]. Subcutaneous implantation of borate bioactive glass with copper microfibers in rats significantly enhanced the growth of capillaries and small blood vessels when compared to silicate 45S5 bioactive glass microfibers [43]. The ionic dissolution product of Cu doped borate bioactive glass microfibers has been shown to stimulate the expression of angiogenic genes of fibroblasts in vitro and angiogenesis in full-thickness skin wounds in rodents in vivo [43]. Moreover Cu doped borosilicate bioactive glass scaffolds have been reported to enhance blood vessel formation and bone regeneration in vivo in rat calvarial defects [25].

Our study has shown for the first time the promising effect of simultaneously using Cu and Sr, elements that, given their great angiogenic and osteogenic potential, were previously independently incorporated in different bioceramics. Further detailed analysis of the ion release from these new MBGs and the study of the long-term influence of the release kinetics on the osteogenic differentiation potential and angiogenic capability of B60, CuSr-1 and CuSr-2.5 glasses will be carried out.

## 5  Conclusions

In this paper, sol-gel bioactive glasses containing up to 2.5% of copper and strontium oxides were synthesized and characterized. The dissolution products of Cu- and Sr- containing bioactive glasses produced the highest secretion of VEGF from ST-2 cells after 48 h of incubation. The combination of an angiogenic stimulator, Cu, and an osteogenic stimulator, Sr, in terms of cell viability and VEGF release has been reported. This combination lays the foundation for engineering highly vascularized, fully functional bone tissue scaffolds for bone regeneration applications.

**Acknowledgements.** This study was supported by research grants from Instituto de Salud Carlos III (PI15/00978) project co-financed with the European Union FEDER funds, the European Research Council (ERC-2015-AdG) Advanced Grant Verdi-Proposal No.694160 and MINECO MAT2015-64831-R project.

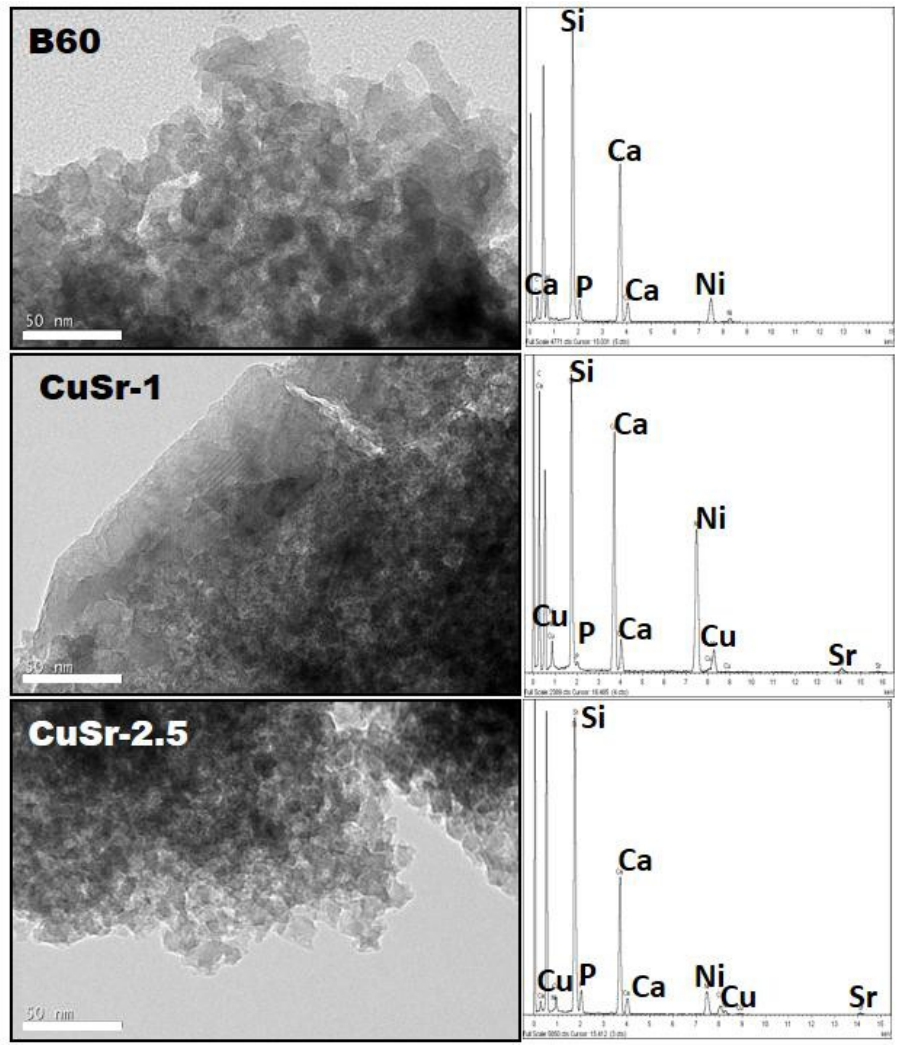

**Figure 1.** TEM micrographs (left) and EDX spectra (right) of B60, CuSr-1 and CuSr-2.5 bioactive glasses.

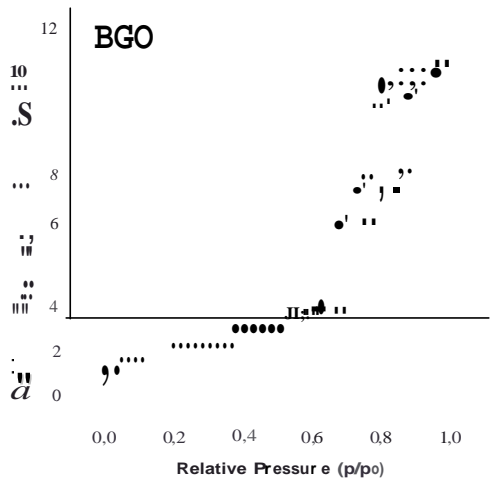
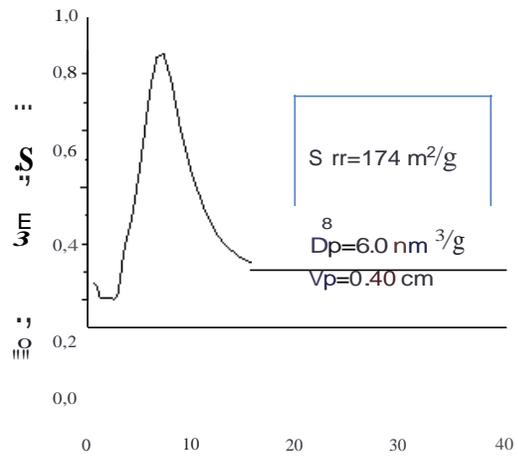
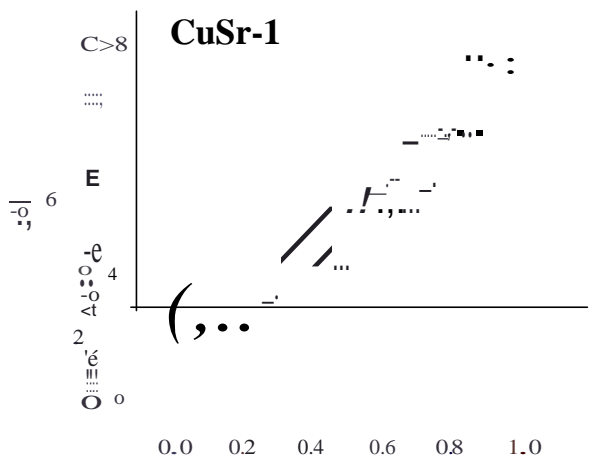
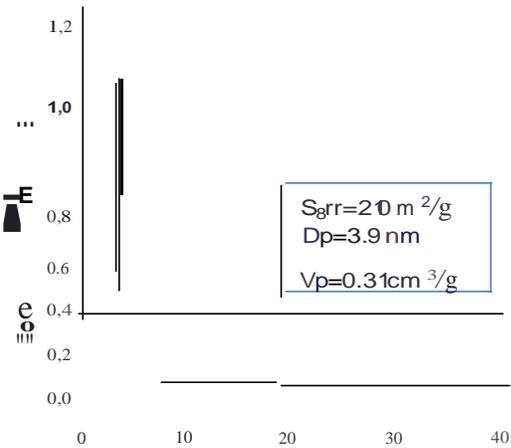
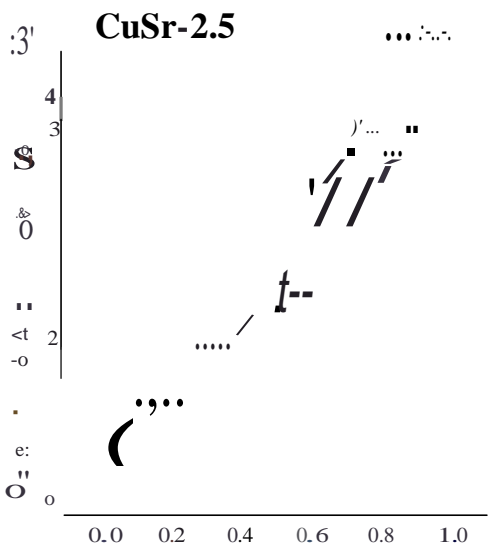
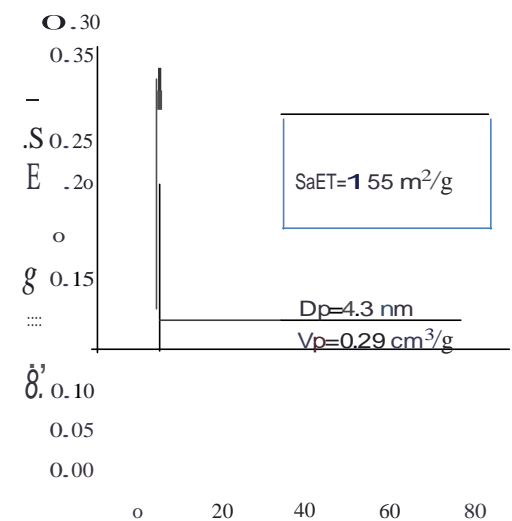

Relative Pressure (p/p°)  Pore {nm}

**Figure 2.** Nitrogen adsorption-desorption isotherms (left) and pore size distributions (right) of B60, OJSr-1 and CuSr-2.5 samples. Inset tables display the most important textural properties of the porous glasses: Surface area (SeEr), pore diameter (Dp) and Volume of pores (Vp).

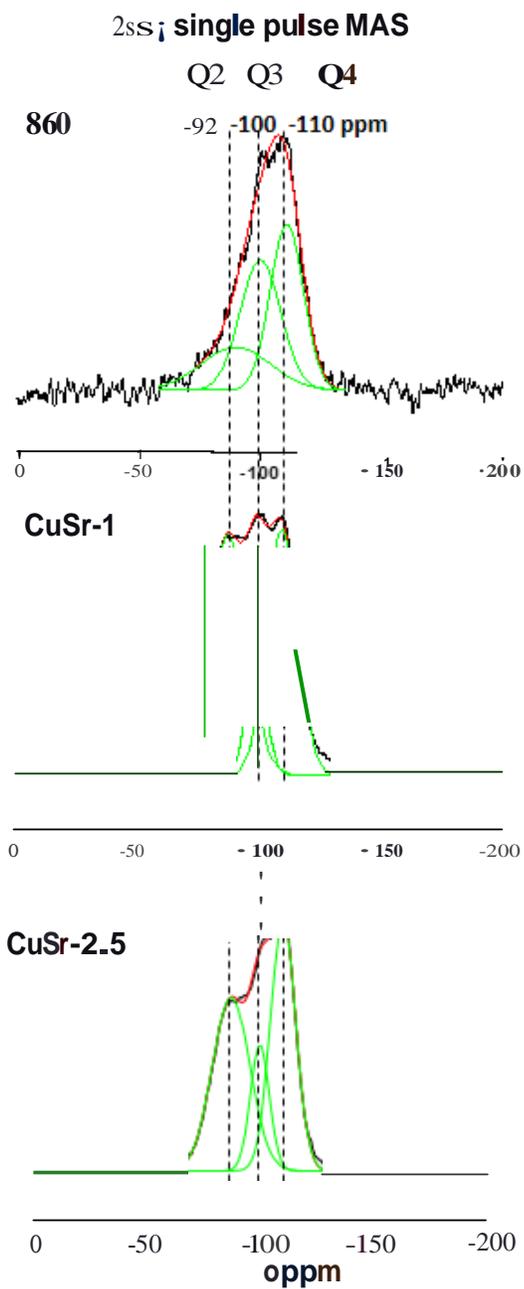

**Figure3.** Solid-state "'Si single-pulse MAS-NMR spectra of B60, CuSr-1 and CuSr-2.5 of B60, CuSr-1 and CuSr-2.S sampies. The areas for the Q" units were calculated by Gaussian line-shape deconvolution and are displayed in green.

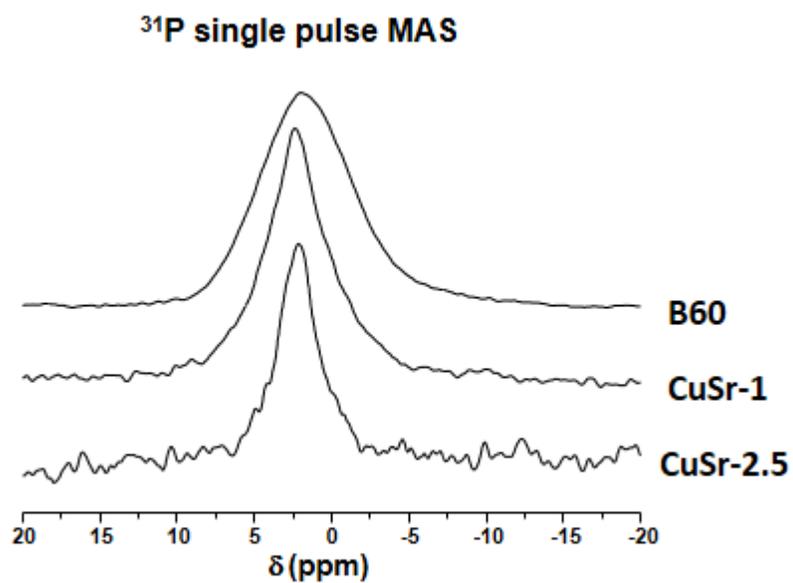

**Figure 4.** Solid-state $^{31}$P single-pulse MAS-NMR spectra of B60, CuSr-1 and CuSr-2.5 of B60, CuSr-1 and CuSr-2.5 samples.

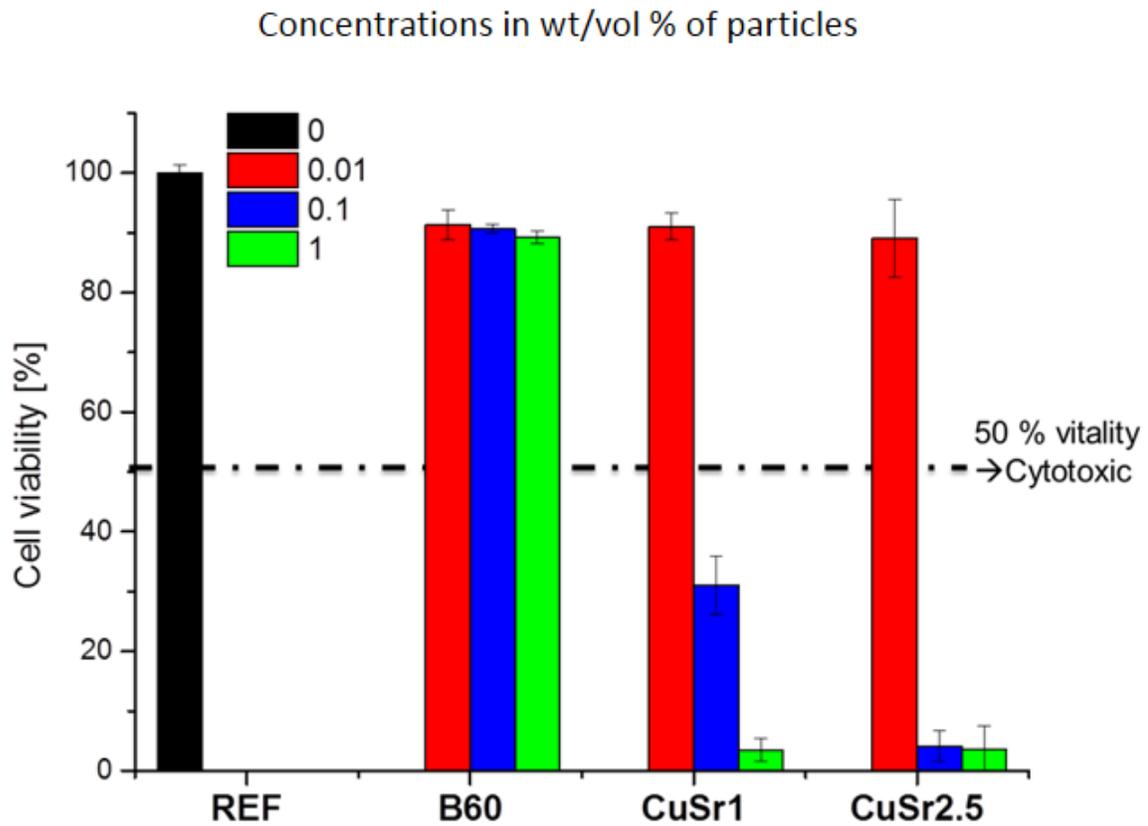

**Figure 5:** Cell viability of ST-2 cells treated with the supernatant of different concentrations (0.01–1 wt/vol %) of B60, CuSr-1 and CuSr-2.5 bioactive glasses after 48 hours of incubation. All results were normalized to 0 wt/vol % ( = 100%).

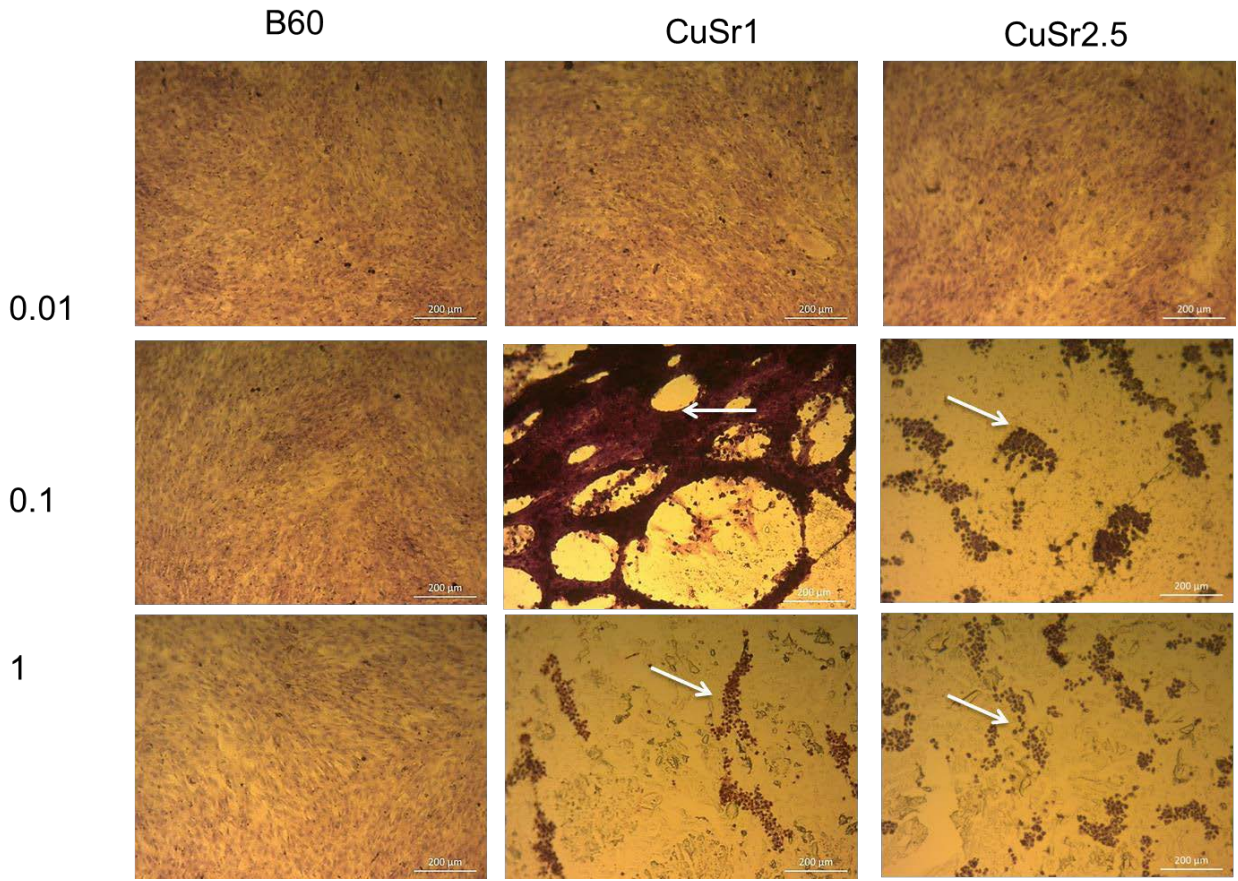

**Figure 6:** Light Microscopy-images of ST-2 cells treated with the supernatant of different concentrations (0.01–1 wt/vol %) of B60, CuSr-1 and CuSr-2.5 bioactive glasses after 48 hours of incubation.

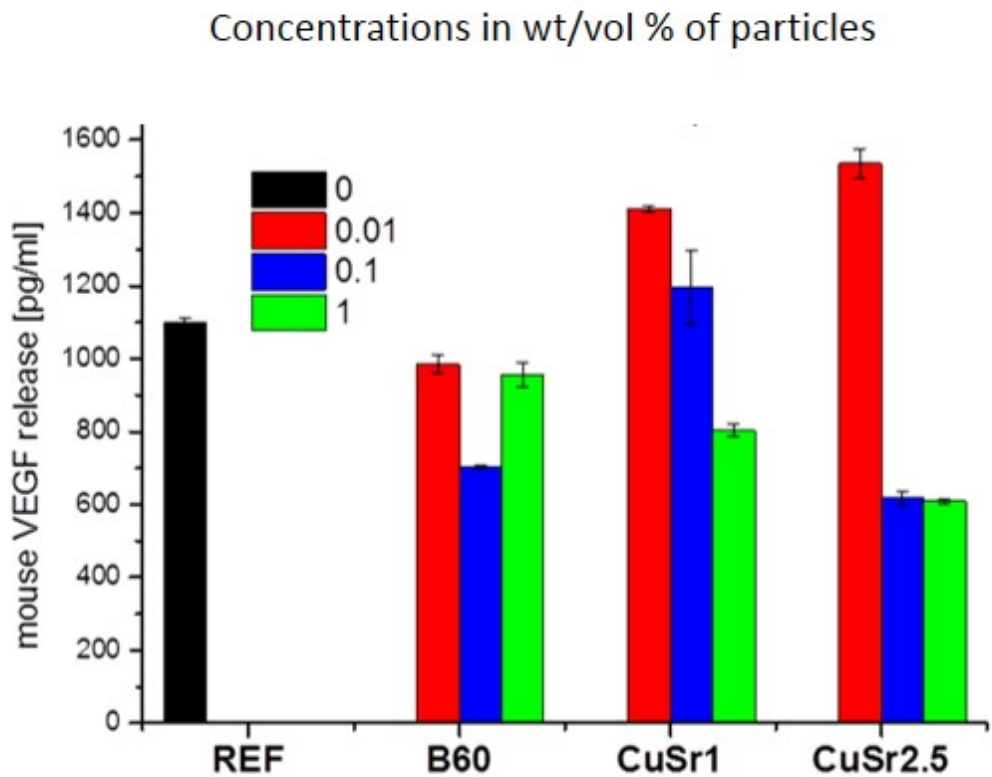

**Figure 7:** Quantification of VEGF released by ST-2 cells after 24 h of incubation with B60, CuSr-1 and CuSr-2.5 particles at different concentrations